\documentclass[manuscript]{aastex}

\slugcomment{Submitted to Ap.J.}

\shorttitle{Viscous-like solar wind-comet tail interaction}
\shortauthors{Reyes-Ruiz, V\'azquez \& P\'erez-de-Tejada}

\begin{document}


\title{Viscous-like Interaction of the Solar Wind \\
       with the Plasma Tail of Comet Swift-Tuttle}


\author{Mauricio Reyes-Ruiz, Roberto Vazquez}
\affil{Instituto de Astronom\'{\i}a,\\
	Universidad Nacional Aut\'onoma de M\'exico,\\
	Apdo. Postal 877,\\
	Ensenada, B.C. 22800, M\'exico \\
	maurey@astrosen.unam.mx}

\and

\author{Hector Perez-de-Tejada}
\affil{Instituto de Geof\'{\i}sica, \\
	Universidad Nacional Aut\'onoma de M\'exico,\\
	Ciudad Univeritaria,\\
	M\'exico D.F. M\'exico}



\begin{abstract}
We compare the results of the numerical simulation of the 
viscous-like interaction of the solar wind with the plasma tail 
of a comet, with velocities of H$_2$O+ ions 
in the tail of comet Swift-Tuttle determined by means 
of spectroscopic, ground based observations. 
Our aim is to constrain the value of the basic parameters 
in the viscous-like interaction model: the effective Reynolds 
number of the flow and the interspecies coupling timescale. 
We find that in our simulations the flow rapidly 
evolves from an arbitrary initial condition to a 
quasi-steady state
for which there is a good agreement 
between the simulated tailward velocity of H$_2$O+ ions
and the kinematics derived 
from the observations. The fiducial case of our model, 
characterized by a low effective Reynolds number
($Re \approx 20$ and 
selected on the basis of a comparison to {\em in situ} 
measurements of the plasma flow at comet Halley, 
yields an excellent fit to the observed kinematics. 
Given the agreement between model and observations, 
with no {\it ad hoc} assumptions, we believe 
that this result suggests that viscous-like momentum transport
may play an important role in the
interaction of the solar wind and the cometary plasma environment.
\end{abstract}



\section{INTRODUCTION}

The interaction of the solar wind with the plasma environment of a comet
has been the subject of numerous studies for the past 50 years. The 
basic model to describe how this interaction takes place, stems  
from the work of Biermann (1951), Alfven (1957), Biermann et al. (1967) and 
Wallis (1973), and involves primarily the effect of mass loading 
of the solar wind by the picked-up, newly born cometary ions 
from the expanding neutral envelope of the comet, and the draping of
the interplanetary magnetic field into its magnetic tail. Cravens and Gombosi 
(2004) and Ip (2004) review the current understanding of the 
interaction of the solar wind with the plasma environment of comets on the 
basis of these processes. 

In addition to these effects, P\'erez-de-Tejada et al. (1980) and
P\'erez-de-Tejada (1989) proposed that, as it happens in 
other ionospheric, unmagnetized obstacles to the solar wind,
such as Venus and Mars, several features of the large scale flow 
dynamics in the plasma environment of comets, can be attributed to 
the action of viscous-like forces as the solar wind interacts with 
cometary plasma. As suggested by Reyes-Ruiz et al. (2009, hereafter 
referred to as RR09), momentum and heat transport coefficients 
in cometosheath plasmas can be greatly increased with respect to 
their normal, ``molecular'' values, owing to the turbulent 
character of the flow (Baker et al. 1986, Scarf et al. 1986, 
Klimov et al. 1986, Tsurutani \& Smith 1986) and/or to 
the development of plasma instabilities leading to effective 
wave-particle interactions, as discussed by 
Shapiro et al. (1995) and  Dobe et al. (1999 and references therein) 
in the ionosheath of Venus.  

{\em In situ} measurements by the Giotto spacecraft
indicate that, as in Venus and Mars, the solar wind flow in the ionosheath 
of comet Halley exhibits an intermediate transition, also called 
the ``mystery transition'', approximately half-way between 
the bow shock and the cometopause (Johnstone et al. 1987, Goldstein et al. 
1986, Reme 1991). Below this transition,
as we approach the cometopause, the antisunward velocity of the shocked solar 
wind decreases in a manner consistent with a viscous boundary layer 
(P\'erez-de-Tejada 1989). Also  indicative of the presence of viscous-like
dissipation processes, 
the temperature of the gas increased and the density decreased, as the
spacecraft moved from the intermediate transition to the cometopause.

In a recent study, following Perez-de-Tejada (1999), RR09 
have argued that the superalfvenic
character of the flow in the cometosheath, downstream from the nucleus and in the
tail region, as found from {\em in situ} measurements at comet Giacobinni-Zinner, 
suggests that viscous-like effects are, at least, as important as ${\bf J} \times {\bf B}$
forces throughout the region. The fact that $M_A^2 >> 1$ in the 
cometosheath means that the magnetic 
energy density is much smaller than the kinetic energy associated with the 
inertia of the plasma. This implies that ${\bf J} \times {\bf B}$ forces
are not the dominant dynamical factor responsible for the large scale 
properties of the flow in the region. By comparing
the magnitude of the terms corresponding to momentum transport 
due to viscous-like forces and ${\bf J} \times {\bf B}$ forces in
the momentum conservation equation, Perez-de-Tejada (1999) 
argued that downstream from the terminator in the ionosheath of Venus, a scenario 
analogous to the one considered in this paper, the fact that the flow is 
superalfvenic, as found from the {\em in situ} measurements of 
the Mariner 5 and Venera 10 spacecraft, indicates that viscous-like forces 
may dominate over ${\bf J} \times {\bf B}$ forces in the flow dynamics
in the boundary layer formed in the interaction of solar wind and 
ionospheric plasma. If the flow is characterized by a low effective Reynolds 
number, $Re$, this layer extends over a significant portion 
of the ionosheath of the solar wind obstacle. RR09 have constrained the value 
of the effective Reynolds number of the flow, $Re$, by comparing the 
results of numerical simulations of the interaction between 
the solar wind and a cometary plasma tail, taking into account the effect
of viscous-like forces, with {\em in situ} measurements
in the ionosheath of comets Halley and Giacobinni-Zinner.
They conclude that a value of $R  \lesssim 50$ is
necessary to reproduce the relative location of the cometopause, intermediate 
transition and bow shock in these comets. 

In this paper we continue our study of the hypothesis that 
viscous-like forces are important in the solar wind-comet plasma 
interaction, by comparing the flow properties along the plasma 
tail of a comet, as derived from the 
results of our numerical simulations, with the kinematics of H$_2$O+ ions 
determined from spectroscopic, ground-based observations of comet 
Swift-Tuttle during late 1992 (Spinrad et al. 1994). 
Since the precise origin of the viscous-like momentum 
transfer processes invoked in our simulations is not yet clear, our study 
will contribute to the
understanding of this process by placing constraints on the parameters, 
such as the effective Reynolds number, that control the flow dynamics. 
This being the first attempt to include viscous-like effects in 
models of the solar wind-comet interaction, several, potentially  
important factors have been neglected, such as the effects of 
magnetic fields, possibly important in the midplane of the plasma 
tail (see above), and ion-neutral collisions, important closer to the comet 
nucleus. These issues are further discussed in RR09 and shall be considered
in future studies.

The paper is organized as follows: in section II we review our numerical 
simulations of the viscous interaction between the solar wind and the 
plasma tail of a comet. In section III we present the observations of 
Spinrad et al. (1994), and the inferred velocity profiles that will 
be used in the comparison. The results of the comparison between model and 
observations are presented in section IV. In section V we discuss several issues 
related to our results, such as time dependence of the flow on the timescale 
of the different observations. Finally, a summary of our results and 
our main conclusions are presented in section VI.

\section{DECRIPTION OF OUR NUMERICAL SIMULATIONS}

Profiles for the tailward velocity of the H$_2$O+ cometary plasma along
the tail,
are obtained from the set of numerical simulations described in detail in
RR09. Briefly, we use a 2D, hydrodynamic, two species,
numerical code that solves the continuity, momentum and energy equations
including the effects of viscous-like forces resulting from turbulence and/or
wave particle interactions in the flow. The interspecies coupling is 
taken from the work of Szego et al (2000).   
The code uses an explicit MacCormack scheme which is 2nd order accurate in space 
and time. In the current implementation of the code we use a non-uniform, cartesian
computational grid. The $x_i$ points are geometrically distributed from 
$x_{min}$ to $x_{max}$ with $nx$ elements. 
The $y_j$ points are equispaced at the initial location 
of the tail (from $y = 0$ to $y = 1$ having 50 gridpoints) 
and geometrically distributed from $y = 1$ to 
$y = y_{max}$. In both series the common ratio is 1.02. 
We do not include the effect of magnetic fields (see Introduction) or a 
source of newborn ions in our simulations other than the assumed inflow in
our boundary condition (see below). 
The equations are written in conservative, adimensional form so that 
the solution is characterized by the value of the adimensional parameters:

\begin{equation}
M_o = \frac{V_o}{C_s} 
\end{equation}

\begin{equation}
Re^a = \frac{\rho_o V_o L}{\mu_o^a},
\end{equation}

\begin{equation}
Re^b = \frac{\rho_o V_o L}{\mu_o^b},
\end{equation}

\begin{equation}
\nu_o = \frac{\nu_{ab} L}{V_o}
\end{equation}

\noindent where $M_o$ is the Mach number for the incident solar wind 
flow defined in terms of $V_o$, the normalization velocity, and $C_s$, 
the sound speed based on the normalization value for the temperature. 
$Re^a$ and $Re^b$ are the effective Reynolds number for solar wind protons 
(species $a$) and cometary ions (species $b$) respectively, 
defined in terms of $\rho_o$,
the normalization density, $\mu_o^a$ and $\mu_o^b$, which are the dynamic viscosity
coefficients for species $a$ and $b$ respectively, and $L$ which is 
the normalization length for the $x$ and $y$ coordinates. Finally, $\nu_{ab}$ is a parameter 
reflecting the inverse of the characteristic timescale for coupling 
between species. In our adimensional equations, the latter appears normalized 
by the crossing timescale for the flow $t_{\rm cross} = L / V_o$, i.e. $\nu_o$ 
is the ratio of the solar wind crossing timescale to the characteristic 
timescale for interspecies coupling. Species $a$ 
in our case corresponds to the proton plasma and species $b$ corresponds
to the plasma of H$_2$O+ ions. These adimensional numbers:
$M_o$, $Re^a$, $Re^b$ and $\nu_o$ are the basic parameters of our model.

The simulations are designed to study the dynamics of solar wind and cometary 
plasma in and around the plasma tail of a comet. In our simulation domain 
the $x$ coordinate is measured along the axis of the plasma tail, increasing
in the antisunward direction. The $y$ coordinate is measured 
perpendicular to the plasma tail of the comet starting from the Sun-comet line. 
The origin of our simulation box is located sufficiently tailward of the 
cometary nucleus to justify our assumption that the flow entering through the 
left hand boundary ($x = 0$) is mostly in the $x$ direction.  

The initial condition for our simulations consists of a dense, slow moving layer 
of cold plasma representing the tail extending over the whole $x$ range of 
the simulation box and contained 
between $y = 0$ and $y = 1$. Both gas species, protons and H$_2$O+ ions, are 
present in the tail with a uniform initial density, although the density of 
H$_2$O+ ions is much greater than that of protons. 
From $y = 1.5$ to $y = y_{\rm max}$ the flow has the properties of the 
shocked solar wind with Mach number $M_o = 2$ in all cases considered for our 
model. Such value
for $M_o$ is taken from the simulations of Spreiter \& Stahara (1980) who
computed the flow in the ionosheath of Venus as the solar wind flows past 
the planet's terminator towards the tail. No significant differences in our 
results are found using a value of $M_o$ between 1.5 and 3. A value of the
Mach number outside this range is not expected. For $y > 1.5$, 
fast moving, hot protons are 
the dominant species with a density 400 times lower than the density of H$_2$O+ ions
in the tail. Between $y = 1$ and $y = 1.5$ there is a transition region 
from the cometary tail flow (below $y = 1$) to the solar wind flow (above $y = 1.5$).

The boundary conditions for our simulations are the following. At the 
left hand boundary we assume a steady inflow preserving the properties 
of the initial 
condition described above. At the top boundary ($y = y_{\rm max}$) we assume 
the flow has the properties of the solar wind flow as in the initial condition.
At the bottom boundary ($y = 0$, the tail midplane) we assume that 
the flow is symmetric, 
so that the $y$ derivatives of all quantities are taken as zero. At the 
right hand boundary we use a zero derivative outflow boundary condition.   

\begin{figure}
\epsscale{.80}
\plotone{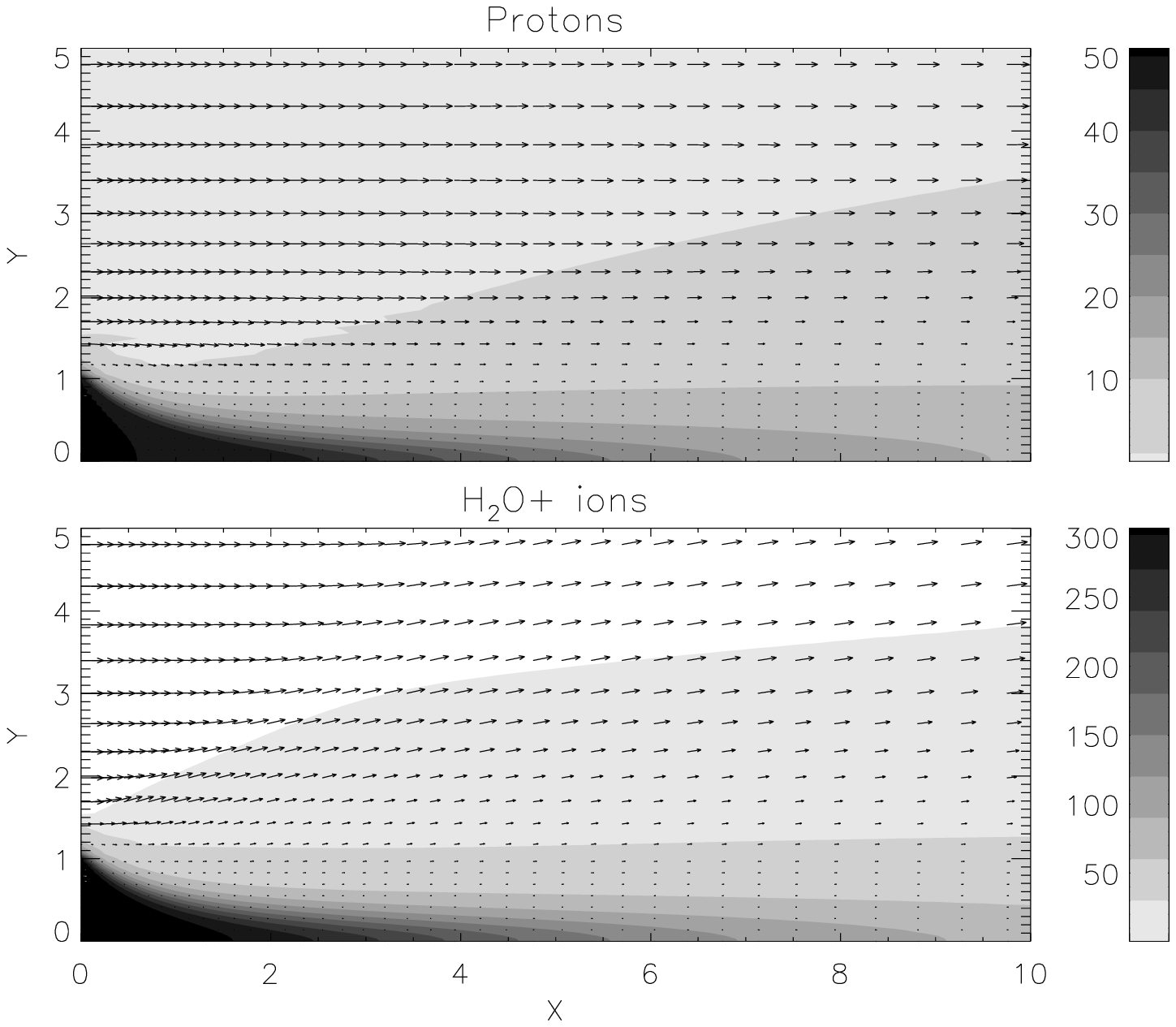}
\caption{Density contours (shades of gray) and flow geometry 
(velocity vectors) for a model characterized by $Re^{a,b} = 30$ 
and $\nu_o = 0.1$ after 1500 simulation time units. 
The top panel shows the configuration for the proton plasma 
(species $a$) and the right side panel shows the ``equilibrium'' 
configuration for cometary H$_2$O+ ions. Density and velocity 
are in normalized units.
\label{fig1}}
\end{figure}

The basic properties of the flow resulting from our simulations are shown 
in Fig. \ref{fig1}. Density contours and velocity vectors for both,
solar wind protons and cometary H$_2$O+ ions, once the flow settles to a 
quiescent evolution, are shown.
A brief description of the main results of RR09 is as follows.
Two processes dominate the dynamics of the flow, first,  
viscous stresses transfer momentum from the solar wind to the cometary plasma in 
the tail giving rise to a boundary layer above the cometopause and accelerating 
cometary material tailwards. Secondly, cometary plasma diffuses into the solar 
wind owing to the relatively weak interspecies coupling. The action of these 
processes results in the development of a distinct transition in the flow,
which can be identified as the height of the viscous boundary layer formed
over the obstacle, intermediate between the shock front and the cometopause. 
The precise location of these transitions depends on the 
model parameters and we have found that a model characterized by 
$Re^a = Re^b = 30$ and $\nu_o = 0.1$, our fiducial model, agrees well with 
the {\em in situ} measurements at comets Halley and Giacobinni-Zinner (RR09).  

As we have stated, the main aim of this paper is to constrain the 
value of the Reynolds numbers and the interspecies coupling timescale in
our models, by comparing the results of simulations 
using different values of these parameters, with the kinematics 
determined from the observations.

\section{OBSERVATIONAL DATA}

On six nights between November 23 and December 24 of 1992,
Spinrad et al. (1994) carried out long-slit spectroscopic observations 
of comet Swift-Tuttle using the
Lick Observatory 0.6 m coud\'e auxiliary telescope with an echelle spectrograph 
in long slit mode and a narrow-band filter centered at $\lambda$6199 {\AA}. 
Several rotational transition lines 
of the H$_2$O+ ion are located in the wavelength range of the filter.
From these observations they derived the flow velocity of the H$_2$O+ 
ions along the center of the tail.        

Plasma velocities are derived from the Doppler shift of H$_2$O+ emission
as one moves along the slit. The slit is aligned along the Sun-comet vector 
and greater shifts are observed as distance from the cometary nucleus 
increases. The slit covers a distance, at the comet, up to 
4 $\times$ 10$^5$ km tailward from the nucleus, depending on the 
comet-Earth distance. 

\begin{figure}
\epsscale{.80}
\plotone{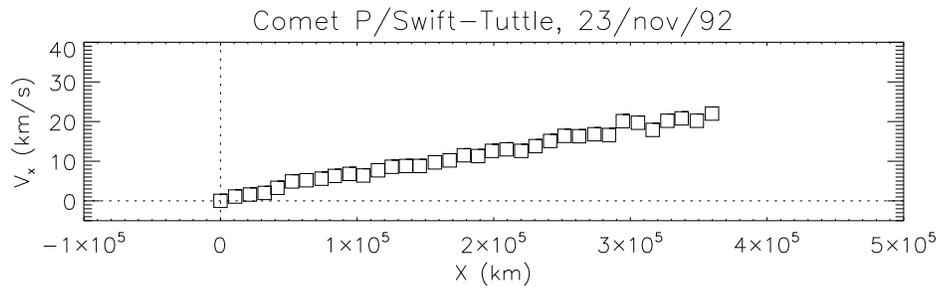}
\caption{Velocity of H$_2$O+ ions in the tail of comet Swift-Tuttle
as derived from the spectroscopic observations by Spinrad et al. (1994).
\label{fig2}}
\end{figure}

Spinrad et al. (1994) found that, in general, velocity increases in a more 
or less linear manner as one moves away from the nucleus along the Sun-comet
vector in the antisunward direction. Typically the velocity increases from 
zero (or almost) at the cometary nucleus to 20-30 km s$^{-1}$ at 3 $\times$ 10$^5$ km 
from the nucleus as illustrated in Fig. \ref{fig2}.  

\section{RESULTS}

We have carried out a series of numerical simulations with different values
for the basic parameters of the problem; $Re^{a,b}$ and $\nu_o$, and in this section
we compare the resulting profiles of tailward velocity, $V_x$, corresponding to a time 
$t \ = \ 1500 \ t_{\rm cross}$ in our simulations, and measured at the 
first active row of the simulation grid, i.e. at the tail midplane with the 
tail kinematics as observed by Spinrad et al. (1994).

Since the flow variables in 
our simulations are in dimensionless form, in order to compare the 
velocity with the observations we must adopt a definite value for the 
normalization lengthscale and velocity. This in turn sets the value for 
the normalization timescale. The normalization speed is set to 
$V_o = 300$ km s$^{-1}$, the speed of the shocked solar as it flows past
the terminator and into the tail in the models of Spreiter \& 
Stahara (1980). To set the normalization lengthscale consider that 
in our simulations, $L$, is approximately the half-thickness of the plasma 
tail. According to the emission intensity measured by Spinrad et al. (1994)
across the tail of comet Swift/Tuttle, most of the emission detected, 
presumably tracing the gas density, is concentrated in a region 
approximately 5 $\times$ 10$^4$ km wide. Hence, we set 
$L =$ 2.5 $\times$ 10$^4$ km in our simulations. As argued before, 
the origin of our simulation box is set far behind the
cometary nucleus to justify assuming that the flow entering through the 
left hand boundary of our box ($x = 0$) is mostly in the $x$ direction. 
In all cases shown here the origin is located a a distance of 1.5 $L$ 
behind the nucleus. This choice is also dictated by the location behind 
the tail where the observed velocity reaches the assumed inflow 
velocity into our simulation box.   

\subsection{Effect of viscosity}

\begin{figure}
\epsscale{.80}
\plotone{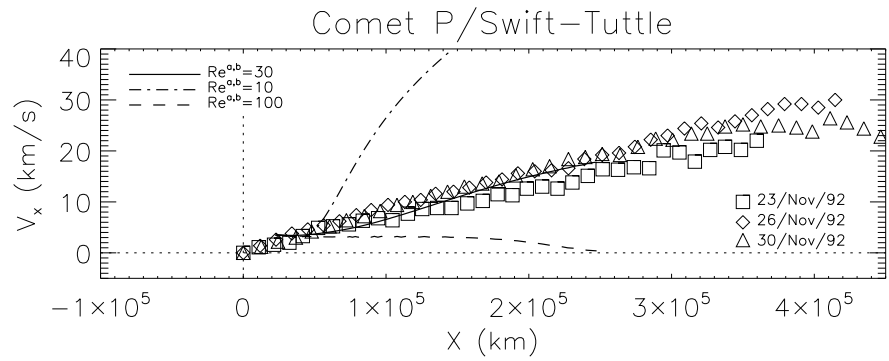}
\caption{Comparison of the tailward velocity of H$_2$O+ ions in our 
numerical simulations (lines) and from the spectroscopic
observations (symbols) of Spinrad et al. (1994) for models with different
Reynolds number. The vertical dotted line 
indicates the position of the nucleus in the observations.
\label{fig3}}
\end{figure}

In Fig. \ref{fig3} we first compare the observations of Spinrad et al. (1994)
with the results of three different cases having different value of the parameter
measuring the relative importance of viscous forces, the Reynolds number, 
$Re^{a,b}$. As discussed in RR09 there are no significant 
differences in the dynamics of the H$_2$O+ ions in cases having different 
values for the Reynolds number for each species, as long as the Reynolds 
number for H$_2$O+ ions, $Re^b$, remains the same. Hence, in Fig. \ref{fig3}
we compare the observed kinematics to the results obtained when
$Re^{a,b} = 10$, $Re^{a,b} = 30$ and  $Re^{a,b} = 100$. All cases are 
characterized by an interspecies coupling parameter $\nu_o = 0.1$. 

As discussed in RR09 the effect of decreasing the
Reynolds number, is to increase the efficiency of momentum transport from 
the solar wind to the cometary material. In doing so, the erosion of
the cometary plasma tail by the solar wind is more effective, and the
plasma in the tail is accelerated to greater velocities as we move along 
the tail. This can be seen from comparing the resulting tailward
velocity of our simulations in Fig. \ref{fig3} for the cases with
$Re^{a,b} = 10$ (dash-dotted line), $Re^{a,b} = 30$ (solid line) and  
$Re^{a,b} = 100$ (dashed line). The case with the lowest Reynolds number 
leads to a much greater acceleration of cometary material while the 
low viscosity case, high Reynolds number, yields almost no acceleration 
of material at the midplane of the tail. 

It is evident in Fig. \ref{fig3} that the case with $Re^{a,b} = 30$
and $\nu_o = 0.1$ (our fiducial case in RR09) gives the 
best fit to the observed plasma velocities and in fact, given their strong
discrepancy with the observations, cases with $Re^{a,b} = 10$ and 
$Re^{a,b} = 100$ can be confidently ruled out. 

\subsection{Effect of interspecies coupling}

\begin{figure}[hpt]
\epsscale{.80}
\plotone{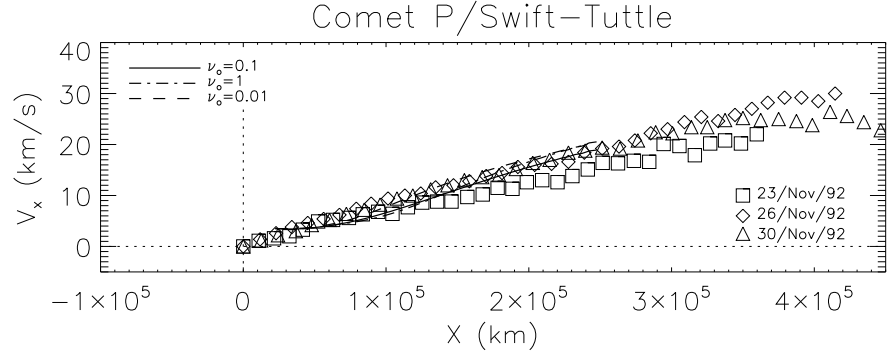}
\caption{Comparison of the tailward velocity of H$_2$O+ ions in 
the numerical simulations (lines) and from the spectroscopic
observations (symbols) of Spinrad et al. (1994) for models with different 
interspecies coupling parameter. The parameter $\nu_o$ is the ratio of the 
the flow crossing timescale $t_{\rm cross} = L/V_o$ to the timescale for
coupling between species. The vertical dotted line 
indicates the position of the nucleus in the observations.
\label{fig4}}
\end{figure}

A comparison of the results of our simulations with the kinematics
determined from the observations of Spinrad et al. (1994) for cases 
having a different interspecies coupling timescale is presented in
Fig. \ref{fig4}. The Reynolds number, $Re^{a,b}$ is the same in all cases.
The effect of decreasing the interspecies coupling is to increase the 
ease with which both species ``penetrate'' each other. 

There are no significant diferences in the H$_2$O+ velocity profiles
along the tail for models with weak ($\nu_o = 0.01$), medium 
($\nu_o = 0.1$) or strong ($\nu_o = 1$) interspecies coupling. 
All three cases agree equally well with the observed kinematics at
comet Swift-Tuttle.   

\section{DISCUSSION}

In this section we seek to discuss the issue of time dependence in 
our solutions and in the observed velocity profiles. 
The results of our numerical simulations are 
evolving constantly, albeit the rate at which the flow properties 
change, slows down considerably after the initial transition from
our arbitrary initial condition. There appears to be also an evolutionary
trend in the kinematics of the flow according in the observations of 
Spinrad et al. (1994). For example, if we compare the observations
of the nights of November 23 and November 26, there is a slight increase 
in the slope of the velocity {\it versus} distance relation, as can be seen from 
comparing the sequence of squares (Nov. 23) and the sequence of diamonds (Nov. 26)
in Fig. 3. Comparing to the velocity corresponding to the night of November 30
(triangles in Fig. 3) there is no appreciable difference to the flow measured 
on November 26. A similar tendency is observed by comparing the 
velocity profiles for the nights of December 23 and 24 in the observations
of Spinrad et al. (1994), albeit these have a shallower slope. 

The results of our simulations are reminiscent of this behavior.  
Considering the values of $L$ and $V_o$ 
adopted, the timescale for the solar wind to cross a distance $L$ is
$t_{\rm cross} \approx 80$ s. Hence, a timescale of 1200  $t_{\rm cross}$ 
in our simulations corresponds approximately to 1 day in the observations
of comet Swift-Tuttle. So, for example, the system takes about 1 day
to go from the full, unperturbed tail scenario of the initial condition, 
to the eroded tail condition illustrated in Fig. 1. As we have argued,
following this rapid evolution process, the flow continues to evolve in a 
quiescent, much slower manner. In Fig. 5 we compare two tailward velocity 
profiles from our numerical simulations taken with a time difference 
of 3 days (solid and dashed lines), with the velocity 
profiles derived from the observations of Spinrad et al. (1994) 
corresponding to the nights of November 23 (squares)
and November 26 (diamonds). If we extend our simulations for another 3 days, 
no significant changes result in the velocity profiles, as the simulations 
have already reached a quasi-stationary state. 
We believe that the difference in flow 
properties observed between the observations of late November and those 
conducted one month later reflect a difference in conditions of the 
solar wind (and IMF) incident on the comet, an effect known to give rise to 
drastic changes in the properties of the plasma tail such as the thickness 
of the tail as it emerges from the nucleus.    

Rather than showing profiles giving an exact fit of our results
to the observations (making the appropriate {\it ad hoc} modifications),
our aim has been to illustrate the fact that the 
evolutionary trend in our model is qualitatively consistent with the 
observations.      

\section{CONCLUSIONS}

In RR09 we have presented the first numerical simulations of the interaction 
of the solar wind with the plasma tail of a comet including the effects 
of viscous-like forces  as originally proposed by Perez-de-Tejada et al. (1980).
In this paper we have carried out a comparison of the tailward velocity
profiles of the cometary ions derived in our numerical simulations,
with the kinematics of H$_2$O+ ions determined from spectroscopic, ground 
based observations along the tail of comet Swift-Tuttle (Spinrad et al. 1994).

Our main result is that the case we have chosen as fiducial in a previous 
study, on the basis of a comparison of our model with {\it in situ} measurements
at comets Halley and Giacobinni-Zinner (RR09), characterized by a low 
Reynolds number ($Re^{a,b} = 30$) and moderate interspecies coupling 
parameter ($\nu_o = 0.1$), gives one of the best fits to the flow 
kinematics observed by Spinrad et al. (1994).  Cases with a
much larger Reynolds number (smaller viscosity) are ruled out by 
the observations. Hence, on the basis of these results, we conclude that 
an efficient viscous-like momentum transfer between the solar wind 
and cometary material in the plasma tail, is necessary to explain 
the observed kinematics in comet Swift-Tuttle. The fact that a similar 
conclusion is reached from the analysis at comets Halley and Giacobinni-Zinner  
(RR09) suggests that viscous-like momentum transfer may be a dynamically 
important process in the interaction of the solar wind and the plasma 
environment of comets in general. 

We finalize by pointing out several issues still to be addressed, 
that may have important consequences on the results of our 
simulations and on the conclusions of this study. Our 
simplified treatment does not consider the ongoing creation of
H$_2$O+ ions in the tail region due to the ionization of cometary
neutrals. At this stage in our ongoing modeling effort, we have also 
neglected the effect of the draped IMF on the flow dynamics (although
as discussed in RR09 we do not expect these to be dominant), 3D effects, 
and time dependence of the incoming flow. The effect of the 
precise form of the viscous-like transport and effective interspecies 
coupling terms in the equations of motion, is another pending issue. We  
believe that the further assessment of the relevance of these factors 
is beyond the present study. They are the subject of work
currently in progress and will be reported in future contributions.
 
\acknowledgments

MRR acknowledges the support of research grant IN10855 of DGAPA-UNAM. The 
authors acknowledge the contribution of students Sergio Dominguez, Adriana
Gonzalez and Rosa Isela Gaspar who participated in the initial stages 
of this work.

\clearpage



\clearpage

\clearpage

\clearpage

\clearpage

\begin{figure}[hpt]
\epsscale{.80}
\plotone{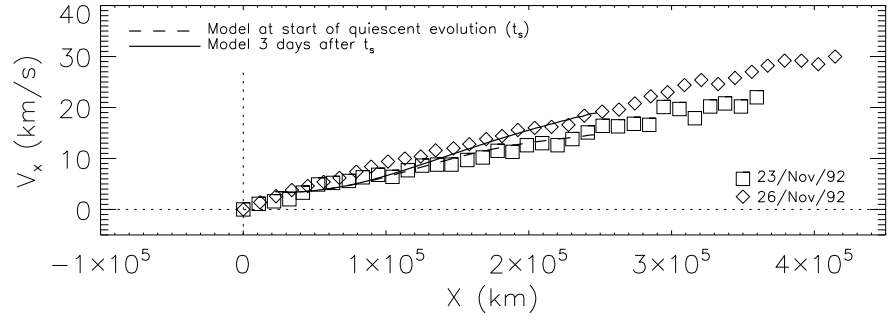}
\caption{Comparison of the evolution of the tailward velocity of 
H$_2$O+ ions in our fiducial model ($Re^{a,b} = 30$, $\nu_o = 0.1$) in 
the numerical simulations (lines), and from the spectroscopic 
observations (symbols) of Spinrad et al. (1994).
\label{fig5}}
\end{figure}


\end{document}